# Role identification of social networkers

Anna Zygmunt

AGH University of Science and Technology, Department of Computer Science

Kraków

Poland

azygmunt@agh.edu.pl

## Synonyms

finding social positions, key users

## Glossary

**Key members (users)** - those who contribute to the success and health of the community

**Social media** - set of web-based technologies targeted at forming and enabling a potentially massive community of participants to productively collaborate

**Categories of social media**: *Blogs* (e.g. Wordpress, Blogcatalog), *Friendship Networks* (e.g. Facebook, MySpace, LinkedIn), *Microblogging* (e.g. Twitter), *Media Sharing* (e.g. (Flickr, YouTube), *Social Bookmarking* (e.g. Del.icio.us), *Social News* (e.g. Digg), *Social Colaboration* (e.g. Wikipedia, Scholarpedia)

**Social networker**: person building her/his position by creating its own social networks in a variety of social media (categories) (http://www.wikihow.com/Be-a-Social-Networker)

## Definition

The concept of "social role" has been the subject of analysis for more than 100 years, which underlines the importance of the problem. The meaning of this term is broad and definitions largely dependent on the application. Therefore, it is difficult to give one definition which would widely be recognized.

Most of the definitions of roles in network analysis have their origin mainly in sociological and psychological research, where social roles are treated as "cultural objects that are recognized, accepted and used to accomplish pragmatic interaction goals in a community" [Gleave 2009]. Sociological research emphasizes the importance of relations to others and expectations for systematic behavior. So roles describe the intersection of behavioral, meaningful and structural attributes that emerge regularly in particular settings and institutions [Welser 2011].

In role theory, roles are defined as "those behaviors characteristic of one or more persons in a context" [Biddle 1986]. In turn, in network analysis: a role is identified as a position that has a distinct pattern of relations to other positions [Wasserman 1994]. In social media, a definition of a role that seems to be most appropriate treats it as a set of characteristics (relevant metrics) that describe behavior of individuals and their interactions between them within a social context [Junquero 2012].

## Introduction

A social network consists of a set of actors and a set of relationships between them which describe certain patterns of communication. Most current networks are huge and difficult to analyze and visualize. One of the methods frequently used to analyze social networks is to extract the most important features, namely to create a certain abstraction, that is the transformation of a large network to a much smaller one, so the latter is a useful summary of the original one, still keeping the most important characteristics. In the case of a social network it can be achieved in two ways. One is to find groups of actors and present only them and relationships between them. The other is to find actors who play similar roles and to construct a smaller network in which the connection between the actors would be replaced with connections between the roles.

Work on assigning actors to roles greatly intensified with the advent of the possibility of collecting vast amounts of data capable of being then analyzed using methods of social network analysis. Social media, whose rapid growth can be observed for several years provide us with new opportunities to define and use roles.

Classifying actors by the roles they are playing in the network can help to understand 'who is who'. This classification can be very useful, because it gives us a comprehensive view of the network and helps to understand how the it is organized, and to predict how it could behave in the case of certain events (internal or external).

In the beginning, the analysis of social media (mainly Usenet - the oldest social media developed in 1979) indicated a very unbalanced participation of users, i.e. most messages were written by a small percentage of users. It was thought that the identification of such popular users would help to understand processes taking place in the community. Many approaches, therefore, focused on finding only *leaders* in the community.

Another group of similar topics is finding *influential users*. Referring to the fundamental article [Keller 2003] attempts were made to translate influential users characteristics (e.g. recognition, generation activity, novelty, eloquence) described therein into SNA measures.

However, to become important or influential, the group must have users performing different roles (more or less important) that will support such key users and influential users, as well as cause the group formed around these users to be more or less stable.

A lot of studies relate to certain social media and attempt to define their specific roles, so that a different set of roles is characteristic of discussion forums, blogosphere, etc. It is based of the assumption that different communities have different needs and the roles that support these needs vary greatly [Nolker 2005]. In addition, even for the same social media, different authors define different roles, different naming and characteristics.

Roles are also considered as a tool for simplifying patterns of action, distinguishing between different types of users and understanding human behavior.

Since it was noticed that human activity is determined and restricted by social structures (such as social context, history of actions, structure of interactions, attributes people bring to the interaction) [Gleave 2009], the concept of "social roles" is being broadened by a combination of social psychology, social structures and behaviors.

Thus, a huge network can be defined as the interaction of relatively small sets of roles that exist in different proportion, making up a diversity of role ecologies.

## Characteristics of social media

People use various social media services for different reasons, in different ways, and generally behave accordingly in each of them. For example, they use Flickr to display photos with friends; Twitter to show their status; Facebook to be in touch with friends, blogs to express their opinions, interests and beliefs; Delicious to bookmark Web pages, etc. [Agarwal 2012].

Members demonstrate different activity levels, but generally only a small percentage of them are active, participating regularly in discussions and with a very large number of followers, as well as the number of friends and social connections. The rest are not very active, which confirms the power law phenomena [Mathioudakis 2009].

Some people are active only in few social media sites, while others use most of the sites. Each social media has a different structure and a different way to use it. However, it must be noted, that the vast majority have the structure containing the link information and content information [Agarwal 2012] and are built around the message thread data structure (some messages are sent as a reply to a particular previous message).

By combining information about the roles performed by users on various social networking sites more extensive (comprehensive) user profile can be obtained. If the user has a similar behavior on various social media, their behavior on other media can predicted, without the need to analyze them. It can be indicated what types of media will inspire greater commitment of users and why. Identification of the same users on different social networking sites is not easy, and often an assumption that users behave similarly performing similar roles is used (e.g. in [Agarwal 2012] they discovered that the user who is influential on one site, has also a tendency to be influential on the other, using similar style of speaking).

## Examples of roles performed by social networkers

A lot of the early studies on finding roles dealt with Usenet discussion groups. After some time, it turned out that some similar roles can be identified in other threaded discussion spaces (e.g., in the majority) where some messages are sent as a reply to a particular previous message, and in this way a chain of messages is created [Welser 2007] [Fisher 2006].

Based on a combination of visualization of authors posting behavior, posting behaviors of each author's neighbors and egocentric network graph, three important roles were identified in such spaces: "answer person", "discussion person" and "reply magnet" .

The primary mode of interaction for the "answer person" is to provide useful answers to other questions asked by members of the group. "Discussion people" reply to one another about the topics introduced by the topic started by "reply magnet". "Discussion people" are characterized by frequent reciprocal exchanges with a relatively high number of other participants. This social role is the source of most of the discussion content contributed to long threaded conversations. "Reply magnet" is responsible for the majority of messages that initiate long threads. The key behavior of these individuals is creating new threads, usually by posting quoted material from external news sources.

It was noted that the two roles: "answer person" and "discussion person" are critical for many threaded discussion media; their presence contributes to the fact that given social media are living: questions are asked, answers given, there are new topics for discussion. Thus, these roles have a positive impact on the community. Unfortunately, negative roles, such as "spammers" or "flammers" may appear in community.

Analyzing Usenet from the perspective of finding important roles for the durability of community, two types of roles were identified: "leaders" (spread knowledge and maintain the cohesiveness of the group) and "motivators" (keep conversation going) [Nolker 2005]. Both roles were defined on the basis of their behavior, conversations and member relationships. The third role, "chatters", was introduced, which refers to those who are engaged in a single discussion but rarely get involved in other discussions.

Defining a role only on the basis of an individual's behavioral patterns, other roles with different characteristics were found [Viegas 2004]: "answer persons (pollinators)", "debaters", "spammers" and "conversationalists". For example, "pollinators" are characterized by a high number of days active, mostly responding to threads started by other authors with one or just a small number of messages sent to each thread. In turn, "debaters" with a high number of days active, mostly respond to threads started by other authors with large numbers of messages sent to each thread.

Recently, many studies concern Twitter (mainly due to the easy availability of data and the possibility of analysis of the network dynamic ). In the literature, many different sets of roles have been proposed, for example: "mainstream news source" (spreads information through the network); "celebrities" (public figures followed by many persons); "opinion leaders" (spread widely their opinions and exercise a big influence among their persons in the network). A negative role is performed by "social spammers" who use social network to disseminate malware, or spread commercial spam messages [Cha 2010] [Junquero 2012]. Another set of roles is presented in [Fazeen 2011]: "leaders" (who start tweeting, but do not follow others, but they can have many followers); "lurkers" (generally inactive, but occasionally follow some tweets); "spammers" (the unwanted tweeters, also called twammers), and "close associates" (including friends, family members, relatives, colleagues, etc.).

In [Welser 2011][Gleave 2009], roles in Wikipedia were identified (noting that Wikipedia differs from discussion spaces in that the primary activity of the community is the construction of an artifact): "technical editors"(correct small errors related to style or formatting of articles); "vandal fighters" (revert vandalism and sanction norm violators); "substantive experts" (improve the quality of the content of the articles); "social networkers" (support community aspects of Wikipedia and contribute little to the content and form of articles directly). As can be seen, there is quite a large flexibility in defining sets of roles.

# The methodology of finding roles

The most general approach to finding roles consists of two main stages [Wesler 2007] [Junquero 2012]: an in-depth understanding of the community in order to identify roles which may be detected, and then creation of a role with observed characteristics and rules that will allow the classification of individuals into the pre-defined roles.

Social roles can be conceptualized at several different levels of abstraction [Gleave 2009] [Welser 2011]. A good starting point is to first identify roles at the level of meaningful social action, and descend to a lower level of abstraction to identify the key behavioral regularities and distinctive positions in social networks. Then make generalizations to abstract theoretical categories that will make it possible to tie these particular roles to general research objectives.

Finding roles should rely on both structural data and detailed qualitative description of the context and meaning of interaction [Gleave 2009]. In the past, this approach was very difficult (only one of these aspects could be taken into consideration). Now, we have great opportunities to find significant structural roles and understand their meaning within the social context.

Social roles are often inherently defined in relational terms: a role only exists in relation to others, who are likewise enacting social roles [Welser 2007]. Therefore, it is necessary to adopt a macro perspective that combines both individual behavior and ecology of the entire social roles (balance and interaction of roles within a given population [Gleave 2009]) within a given social space.

## Methods of finding social roles

There are four main approaches to identifying social roles.

### *Approach based on equivalence classes*

It is the oldest approach to identifying roles [Wasserman 1994]. It is assumed that a given person is a representative of a group of people who are somehow "equivalent"; that are similar to and different from those of the other categories. Categories are defined in the context of the similarity between patterns of relationships among actors. Formally, these categories are presented using a family of algebraic equivalence classes on nodes: structural, automorphic and regular and their variants, and with the use of blockmodels which are produced by applying these reduction to social networks. The least restrictive reduction is regular equivalence and is best suited for the sociological concept of the role: two nodes are said to be regularly equivalent if they have the same profile of ties with members of other sets of actors that are also regularly equivalent [Hannemann 2005][Lerner 2005]. Regular equivalence can be treated as clustering actors, according to their social positions [Brynielsson 2012].

There are many algorithms for finding regular equivalence. All are based on searching the neighbourhood of actors to find actors of other types. As long as they have similar „types" of actors at similar distances in their k-neighbourhoods, they are regularly equivalent.

### *Approach based on the identification of the core /periphery structure*

This approach is based on extracting certain areas of varying activities and assigning roles to users based on membership of a particular area.

The concept of the core and the periphery was first introduced in [Borgatti 1999]. In the directed graph two classes of nodes can be distinguished: belonging to a coherent subgraph (the core), in which nodes are connected to each other in some maximal way, and loosely connected nodes (the periphery). The core should have a lot of links with the periphery, and should be connected with the core members much more than with the periphery. In turn, the periphery should indicate mainly nodes in the core, and those in the periphery, but only to a small extent.

A core/periphery model was used in the analysis of dynamics of community (Yahoo!) [Backstrom 2008]. The concept of the core was redefined by introducing a definition of the k-core. Members belonging to the k-core must have an appropriate activity (e.g. replied to and been replied by minimum number of distinct users) during a specified time period. People who do not belong to the k-core are "light" users. Due to the time they are part of, the users can further be classified as "long-core" and „short-core". It was noticed, for example, that it is more likely that the users will be long-core, if they belong to a smaller number of groups (probably because they can then focus on those groups, at the level necessary to become long-term users). Moreover, it was noticed that the core is approximately the same size regardless of the size of the group. The relatively simple model was proposed, which did not take into account the quality of written posts by Yahoo! users (they might as well be spam).

In the case of blogosphere, work on the extracting structure similar to the core /periphery has gone in the direction of discovery A-List blogs, defined as „those that are most widely read, cited in the mass media, and receive the most inbound links from other blogs" [Obradovic2009]. Blogs from A-List are strongly connected with one another, but poorly with the rest of blogosphere, which is a decisive majority (long tail). Blogs from A-list are often linked to from the long tail, often link to each other and rarely link to the long tail. To find blogs from the A-list, the modified definition of the k-core as connected components that contain only nodes of a minimum degree of k was used.

Analyzing the dynamics of social media (Flickr, Yahoo!) it was noted [Kumar 2006] that the structure of the periphery is not homogenous and can be distinguished in the "singletons" (nodes not linked to any other network nodes) and "middle region" (isolated groups who interact with one another, but not with the network as such). These isolated groups are in fact influential individualities that act as "stars" combined with a varying number of other users who, in turn, have very few other connections. An analysis of network dynamics shows that such "stars" may be included in the core, or disappear as soon as they lose interest in the group.

### *Approach based on the analysis of basic SNA measures*

This approach takes into consideration the fundamental features of users' structural position such as their number of neighbors and relies on various centrality measures, i.e. local (within social communities) or global (for the entire network) structural features [Newman 2004][Zygmunt 2012].

There are three traditional roles based on node network structure: "hubs", "brokers" ("gatekeepers") and "bridges" ("pustakers") [Nolker 2005][Denning 2004]. "Hub" is a person who links to many others; "broker" is the only connection between communities; "bridge" links several communities. Such roles can be found by analyzing the basic SNA centrality measures, such as *degree centrality* (the number of conversations that a user is engaged in or the number of users that a user has conversed with); *betweenness centrality* (the number of pairs of other members who can converse with each other directly through a user with shortest distance); *closeness centrality* (average conversation distance between a user and all the others in the community). In [Goldenberg 2009] "hub" was defined as "people with both in- and out-degrees that are larger than three standard deviations above the mean". The presence of such roles in the community promotes the spread of innovations such as: innovations are most likely to spread if "hubs" adopt and recommend them; "brokers" and "bridges" play important roles in spreading the idea to new groups.

An analysis of the basic measures of SNA has been used in several studies to define social roles of "starters" and "followers" on discussion forums and in blogosphere [Hansen 2010] [Mathioudakis 2009] [Sun 2012]. "Starters" receive messages mostly from people who are well-connected to each other, and therefore can be identified by low in-degree, high out-degree and high clustering coefficient in the graph. The distinction between the roles is obtained by combining the difference between the number of in-links and out-links of their blogs.

In a similar way an "answer person" was identified " on discussion forums [Fisher 2006], noting that it is a person who responds to many other people, but rarely to those who provide answers to the questions raised by the community. So it should have high out-degree and low in-degree. For each author one-degree and two-degree egocentric social networks were constructed through patterns of reply. These networks were then grouped (e.g. on the basis of collective in-degree and out-degree, degree distribution coefficient across groups), and for each distribution of the out-degree of each actor's out-neighbors were calculated.

[Welser 2007] observed that social roles can be described using patterned characteristics of communication between network members, which is called "structural signatures". It was hypothesized that it is possible to recognize the roles that people play by measuring behavioral and structural signature of their participation. The goal is to identify general structural features that are associated with a particular role. Ego-centric network and visualization was used to identify structural attributes associated with the role. Thus, for example, "answer persons" are mainly connected with users with low degree, their local networks tend to have small proportions of three cycles (i.e. their neighbors are not neighbors of each other, they seldom send multiple messages to the same recipient (few intense ties), they tend to reply to discussion threads initiated by others, and generally contribute one or two messages per thread. Ego network for an "answer person" is similar to a star and differs from ego network persons performing other two roles. [Gleave 2009]

In [Nolker 2005], to define roles a combination of SNA measures with behavior-based measures from the information retrieval (term frequency and inverse document frequency : TF * IDF) was used. TF*IDF indicates the weight of conversation relationships between members. It was observed that *betweenness* plays an important role in predicting the relationship potential that a member has with other members. In turn, high *in-closeness* indicates a member who provides consistency and high *out-closeness* - a member who spreads knowledge.

### *Approach based on clustering feature vectors*

In this approach, each person in the network is represented by a vector of some of the features that represents its behavior and relationships with the other members of the community. These features can be for example: the number of peoples the user knows, the number of people that know the user, the number of reciprocal relationships of the user, the number of messages that the user receives, and the number of documents that depict the user etc.

Such vectors can then be clustered so that people with similar characteristics are placed in one group, [Maia 2008] [Junquero 2012] [Pal 2011]. Mostly well known in statistics and data mining, *k-means* algorithm was used. The cluster is described with the use of relevant metrics that are important for a given role. Thus, each role can have a different number of relevant metrics. On Twitter, for example, the role of "celebrities" means the most followed and mentioned persons, most connected but generally not the most influential, so the relevant metrics for this role are shown in the number of followers and the number of documents depicting a given person. In turn, the role of "information propagators" is mainly a source of information, so the relevant metrics are expressed in the number of followers, average and maximum information propagation, the number of publications, the number of words in tweets that exist in dictionary. It is the user, who on the basis of the results of the clustering can create a set of roles, therefore, such an approach can be regarded as most universal and one of the few independent of the particular application [Junquero 2012].

# Key Applications.

- Marketing, opinion diffusion, advertising: finding people that will ensure that information about new product will come down to the largest number of other people and will cause, for example, an increase in sales.
- Recommendation systems: individuals, who share the same social role might be expected to share the same taste.
- Political campaign: which roles are necessary to ensure the success of campaign.
- Detecting important members of criminal or terrorist groups.

# Future Directions.

- Searching for new roles, creating a uniform formal model describing the community, taking into account the roles and interactions between them.
- Searching roles in several heterogeneous networks, trying to find roles for a user in multiple, heterogeneous networks.
- Developing recommendations for the structure of the system of roles (roles ecologies) (qualitative and quantitative) for the effective functioning of communities.

# Cross-References.

- Analysis Specific Social Media, 00226
- Centrality Measures, 00227
- Communities Discovery and Analysis in Online and Offline Social Networks, 00006
- Patterns in Productive Online Networks: Roles, Interactions and Communications, 00334
- Role Discovery, 00288

# References.

# Recommended Reading